\documentclass[aps,prb,reprint,letterpaper,showpacs]{revtex4-1}
\usepackage{graphicx}
\usepackage{bm}
\usepackage[colorlinks,urlcolor=blue,anchorcolor=blue,linkcolor=blue,citecolor=blue,breaklinks=true]{hyperref}

\begin{document}
\title{Variational Monte Carlo study of spin dynamics in underdoped cuprates}
\date{\today }

\begin{abstract}
	The hour-glass-like dispersion of spin excitations is a common feature of underdoped cuprates. It was qualitatively explained by the random phase approximation based on various ordered states with some phenomenological parameters; however, its origin remains elusive. Here, we present a numerical study of spin dynamics in the $t$-$J$ model using the variational Monte Carlo method. This parameter-free method satisfies the no double-occupancy constraint of the model and thus provides a better evaluation on the spin dynamics with respect to various mean-field trial states. We conclude that the lower branch of the hour-glass dispersion is a collective mode and the upper branch is more likely the consequence of the stripe state than the other candidates.
\end{abstract}

\author{Zuo-Dong Yu$^{1}$}
\author{Yuan Zhou$^{1,2}$}
\email{zhouyuan@nju.edu.cn}
\author{Chang-De Gong$^{3,1,4}$}
\affiliation{$^1$National Laboratory of Solid State Microstructure, Department of
Physics, Nanjing University, Nanjing 210093, China\\
$^{2}$Condensed Matter Physics and Material Science Department, Brookhaven
National Laboratory, Upton, New York 11973, USA\\
$^{3}$Center for Statistical and Theoretical Condensed Matter
Physics, Zhejiang Normal University, Jinhua 321004, China\\
$^{4}$ Collaborative Innovation Center of Advanced Microstructures, Nanjing University, Nanjing 210093, China\\}
\maketitle

\section{Introduction}
Extensive efforts have been attempted to understand high-temperature superconductivity in cuprates for decades\cite{lee2006}. One of the widely concerned questions is the normal state in the underdoped regime. The $t$-$J$ model, derived from the Hubbard model in the strong correlation limit, is considered to be an effective low-energy model for cuprates\cite{zhang1988}. Within this framework, many ordered states have been proposed to be the ground state, e.g., the spin-density wave (SDW)\cite{moon2009}, d-density-wave (DDW)\cite{chakravarty2001}, and resonating valence bond (RVB) states\cite{yang2006}. Recently, a charge order has been found in various cuprates\cite{fujita2014,comin2015,da_silva_neto2015,fujita2014_1,hamidian2015,wu2015}, suggesting its universality. However, it is hard to determine which state is actually achieved in cuprates by the single-particle properties, such as the doping evolution of the Fermi surface topology and the quasiparticle dispersion. Therefore, some dynamical correlations are suggested to provide further justification. Among them, dynamical spin correlation has been widely studied due to its direct connection with inelastic neutron scattering (INS) measurements.

So far, extensive INS measurements have been performed on various types of cuprates\cite{fujita2011,armitage2010,tranquada2014}. They share some common features, namely the so called hour-glass dispersion with a structural transition of the incommensurability between the lower and upper branch of the dispersion: The lower branch has dominant peaks along the vertical direction at $(\pi,\pi\pm\delta)$ and $(\pi\pm\delta,\pi)$ in the momentum space, forming the diamond shape. By contrast, the square shaped incommensurability with peaks at $(\pi+\delta,\pi+\delta)$ (and its equivalent points) are found in the upper branch. At the neck, the excitation is commensurate at $(\pi, \pi)$, resulting in a resonance at the characteristic energy $E_{res}$. Despite those similarities, the differences between various types of cuprates should be noted as follows. In optimally doped YBa$_{2}$Cu$_{3}$O$_{6+x}$, the hour-glass spectrum is evident in the superconducting state with the commensurate resonance at $E_{res}$ and the second incommensurate resonance at slightly higher energy\cite{bourges2000,pailhes2004,dahm2009}. Whereas, it changes little in La$_{1-x}$Sr$_{x}$CuO$_{4}$ and La$_{2-x}$Ba$_{x}$CuO$_{4}$ below and above the superconducting critical temperature $T_{c}$. For La$_{1-x}$Sr$_{x}$CuO$_{4}$, the hour-glass spectrum persists above $T_{c}$ even at optimal doping and there is no commensurate resonance\cite{christensen2004,vignolle2007}. For La$_{2-x}$Ba$_{x}$CuO$_{4}$, the hour-glass spectrum is found in the normal state with the static stripe order; neither the resonance nor the spin gap exists below $T_{c}$ in the underdoped region\cite{xu2014,tranquada2004} . We would like to point out that the resonance discussed here differs from the superconducting resonance mentioned in many literatures, where it is defined by the difference of spin susceptibility between the superconducting and normal states and follows the simple scaling rule $E_{res}/\Delta_{H}=2$ with $\Delta_{H}$ being the magnitude of the superconducting gap at the hot spot\cite{zhou2013_1}.

On the other hand, the theoretical studies devoted to understand the nature of spin dynamics in the cuprates are generally divided into two categories: one is the itinerant-electron picture based on the Fermi surface topology, which is then treated within the random phase approximation (RPA) on top of either the uniform\cite{brinckmann1999,li2002,das2012,zhang2013,eremin2005,onufrieva2002,norman2007} or striped mean-field orders\cite{seibold2006,andersen2005,christensen2016}. Some considerable successes, for example, the structure transition in the magnetic excitations and resonance features, have been accomplished in the RPA framework. However, the RPA treatment usually contains some phenomenological parameters in order to generate the expected results, making the results somewhat uncertain. Furthermore, this picture does not satisfy the local-spin sum rule and therefore does not respect the no double-occupancy constraint of the $t$-$J$ model. The other category is the localized-spin picture, where the dynamics of the doped holes is neglected by assuming that the holes form into stripes, such as the spin-wave theory\cite{kruger2003,carlson2004} and a coupled two-leg ladders model\cite{uhrig2004,vojta2004}. This approach overemphasizes the bosonic mode and ignores the fermionic nature of the system. A more rigorous method for studying spin dynamics in the cuprates is therefore desirable.

In this paper, we numerically study the spin dynamics of the $t$-$J$ model using the variational Monte Carlo (VMC) method. This approach contains both the localized-spin and itinerant-electron properties after performing locally the no double-occupancy projection on the electronic mean-field wavefunction in the VMC framework exactly. It is parameter free, and therefore provides a better evaluation on the spin dynamics for several widely used mean-field trial states for the $t$-$J$ model. The VMC method used to calculate the dynamical spin correlation is briefly introduced in Sec.~\ref{s1}. The spin spectra are studied based on some widely proposed trial states in Sec.~\ref{s2}, together with some further discussions. The paper is summarized in Sec.~\ref{s3}.

\section{model and methods}
\label{s1}
The model we adopted is the $t$-$t'$-$J$ Hamiltonian defined on the square lattice as
\begin{eqnarray}
	H=&-&t\sum_{\langle i,j\rangle}P_{g}(c_{i\sigma}^{\dagger}c_{j\sigma}+H.c.)P_{g}\nonumber\\
	&-&t'\sum_{\langle\langle i,j\rangle\rangle}P_{g}(c_{i\sigma}^{\dagger}c_{j\sigma}+H.c.)P_{g}+J\sum_{\langle i,j\rangle}S_{i}\cdot S_{j},
\end{eqnarray}
where the Gutzwiller projection operator $P_{g}=\prod_{i}(1-n_{i\uparrow}n_{i\downarrow})$ projects out the double occupancy in Hilbert space. $c_{i\sigma}^{\dagger}$, and $c_{i\sigma}$ are the electron creation, and annihilation operators at the $i$th site with spin $\sigma$, respectively. $t$ and $t'$ are the hopping integrals for the nearest-neighbor and next-nearest-neighbor bonds and $J$ is the antiferromagnetic superexchange coupling constant between the nearest-neighbor spins.

The details of VMC method can be found in many literatures (see, for example, Ref.~\onlinecite{ruger2013,gros1989}). Here we schematically present the main idea and some improvements in calculation of the spin dynamics. It is convenient to write down the trial wave function directly in the real space by using Bogoliubov-de Gennes mean-field Hamiltonian in the case of translational symmetry breaking state. To better account for the superconductivity, we apply the ``partial" particle-hole transformation on spin down electrons by $c_{\downarrow}^{\dagger} \rightarrow h_{\downarrow}$ in hole representation. The mean-field Hamiltonian is
\begin{eqnarray} H^{MF}=&-&\sum_{i,j}(\tilde{t}_{ij}+iD_{ij})(c_{i\uparrow}^{\dagger}c_{j\uparrow}-h_{i\downarrow}^{\dagger}h_{j\downarrow})+H.c.\nonumber\\
	&+&\sum_{\langle i,j\rangle}\Delta_{ij}(c_{i\uparrow}^{\dagger}h_{j\downarrow}+c_{j\uparrow}^{\dagger}h_{i\downarrow})+H.c.\nonumber\\
	&+&\sum_{i}\frac{m_{i}}{2}(c_{i\uparrow}^{\dagger}c_{i\uparrow}+h_{i\downarrow}^{\dagger}h_{i\downarrow})\nonumber\\
	&+&\sum_{i}(n_{i}-\mu)(c_{i\uparrow}^{\dagger}c_{i\uparrow}-h_{i\downarrow}^{\dagger}h_{i\downarrow}).
\end{eqnarray}
Here $\tilde{t}_{ij}=t_{ij}+dt_{ij}$, $D_{ij}$, $\Delta_{ij}$, $m_{i}$, $n_{i}$ and $\mu$ are variational parameters for hopping, d-density wave, superconducting, magnetic, on-site charge density order and chemical potential, respectively. These parameters are used to minimize the ground-state energy of the $t$-$J$ model.

The variational ground state is constructed as
\begin{equation}
\vert G \rangle=P_{N}P_{g}\vert MF\rangle=P_{N}P_{g}\prod_{n,\varepsilon_{n}<0}\sum_{i}U_{n,i}(c_{i,\uparrow}^{\dagger}+h_{i,\downarrow}^{\dagger})|0\rangle,
\end{equation}
where the projection operator $P_{N}$ preserves the particle conservation and $U_{n,i}$ is the unitary matrix for Bogoliubov transformation. $\vert MF\rangle=\prod_{n,\varepsilon_{n}<0}\gamma^{\dagger}_{n}\vert 0\rangle$ is the mean-field ground state, where $\gamma^{\dagger}_{n}=\sum_{i}U_{n,i}(c_{i,\uparrow}^{\dagger}+h_{i,\downarrow}^{\dagger})$ creates the quasiparticle with energy $\varepsilon_{n}$.

The spin excitations measured by INS is directly characterized by the imaginary part of the transverse spin susceptibility $\chi^{-+}(q,\omega)$. Its Lehnmann representation is written as
\begin{equation}
	\Im\chi^{-+}(q,\omega)=\frac{1}{\langle G\vert G\rangle}\sum_{n}\vert\langle n \vert S_{q}^{+}\vert G\rangle \vert^{2}\delta(\omega -(E_{n}-E_{g})).
\end{equation}
Here $\vert n\rangle$ is the excited state with energy $E_{n}$ and $E_{g}$ is the ground state energy. Because the spin operator commutes with Gutzwiller operator $P_{g}$, we have
\begin{equation}
	S_{q}^{+}\vert G\rangle=\sum_{k}\phi_{k}^{0}\vert k\rangle,
\end{equation}
with $\vert k\rangle=P_{g}\gamma_{k+q}^{\dagger}\gamma_{k}^{\dagger}\vert MF\rangle$ and $\phi_{k}^{0}=U_{k+q}U_{k}$. We therefore define a variational space with basis $\vert k\rangle$, which is a total spin-$1$ state. Similarly, the excited state is expressed as $\vert n\rangle=\sum_{k}\phi_{k}^{n}\vert k\rangle$ with $\phi_{k}^{n}$ the eigenvector.

Now we construct the excited state in this variational space by solving the generalized eigenvalue problem of Hamiltonian as
\begin{equation}
	\sum_{k'}\langle k \vert H  \vert k'\rangle\phi_{k'}^{n}=E_{n}\sum_{k'}\langle k \vert k'\rangle\phi_{k'}^{n},
	\label{eqeg}
\end{equation}
where $H_{k,k'}=\langle k \vert H \vert k'\rangle$ is the Hamiltonian matrix and $N_{k,k'}=\langle k \vert k'\rangle$ is the overlap matrix since the basis is not guaranteed to be orthonormal. The matrix elements $H_{k,k'}$ and $N_{k,k'}$ can be calculated by the standard Monte Carlo procedure based on the probability $\frac{\vert \langle G'\vert  i\rangle\vert ^{2}}{\langle G'\vert G'\rangle}$, where $\vert G'\rangle$ is an assuming spin-1 state. In principle, $\vert G'\rangle$ can be chosen arbitrarily. Here, we use one single Monte Carlo procedure based on the probability distribution $\langle G'\vert G'\rangle=\sum_{k}\langle k\vert k\rangle$ to make the procedure more efficient\cite{li2010}.

The final expression for the transverse spin susceptibility is
\begin{equation}
	\Im\chi^{-+}(q,\omega)=\frac{1}{\langle G\vert G\rangle}\sum_{n,k,k'}|\phi_{k}^{n*}\phi_{k'}^{0}N_{k,k'}\vert^{2}\delta(\omega -(E_{n}-E_{g})).
\end{equation}
In practice, we replace function $\delta(x)$ with $\frac{\Gamma}{\pi(x^{2}+\Gamma^{2})}$ with $\Gamma$ the energy broadening. The present scheme satisfies the sum rule in the sense of the excitation space $\vert k\rangle$ including all states produced by $S_{q}^{+}\vert G\rangle$ and therefore respects the no double-occupancy constraint. At half-filling, the approach naturally reproduces the spin-wave excitations expected in Heisenberg model\cite{dalla_piazza2014} by assuming SDW+RVB state. Similar approach had also been used to calculate the single particle spectral function\cite{tan2008}. In order to guarantee the close-shell condition, we adjust the doping level or the boundary condition to remove the ambiguity induced by the degeneracy of the trial wave function.

\section{Results and Discussion}
\label{s2}
We study the spin dynamics in three widely proposed states: the RVB state, DDW state and stripe state. These states are often believed to be the potential candidates for the pseudogap and qualitatively account for the hour-glass feature of the magnetic excitations in cuprates within the RPA framework. We set the model parameters as $t'=-0.3t$, $J=0.3t$ with $t$ taken as unit. The size of lattice is as large as $20\times 20$ with the periodic-boundary condition in the RVB and DDW state while it is $16\times 16$ with the antiperiodic along x direction in the stripe state. The energy broadening is fixed at $\Gamma=0.02$ unless specified.

\subsection{RVB and DDW state}

\begin{figure}[ht]
	\centering
	\includegraphics[width=0.45\columnwidth,clip=true,angle=0]{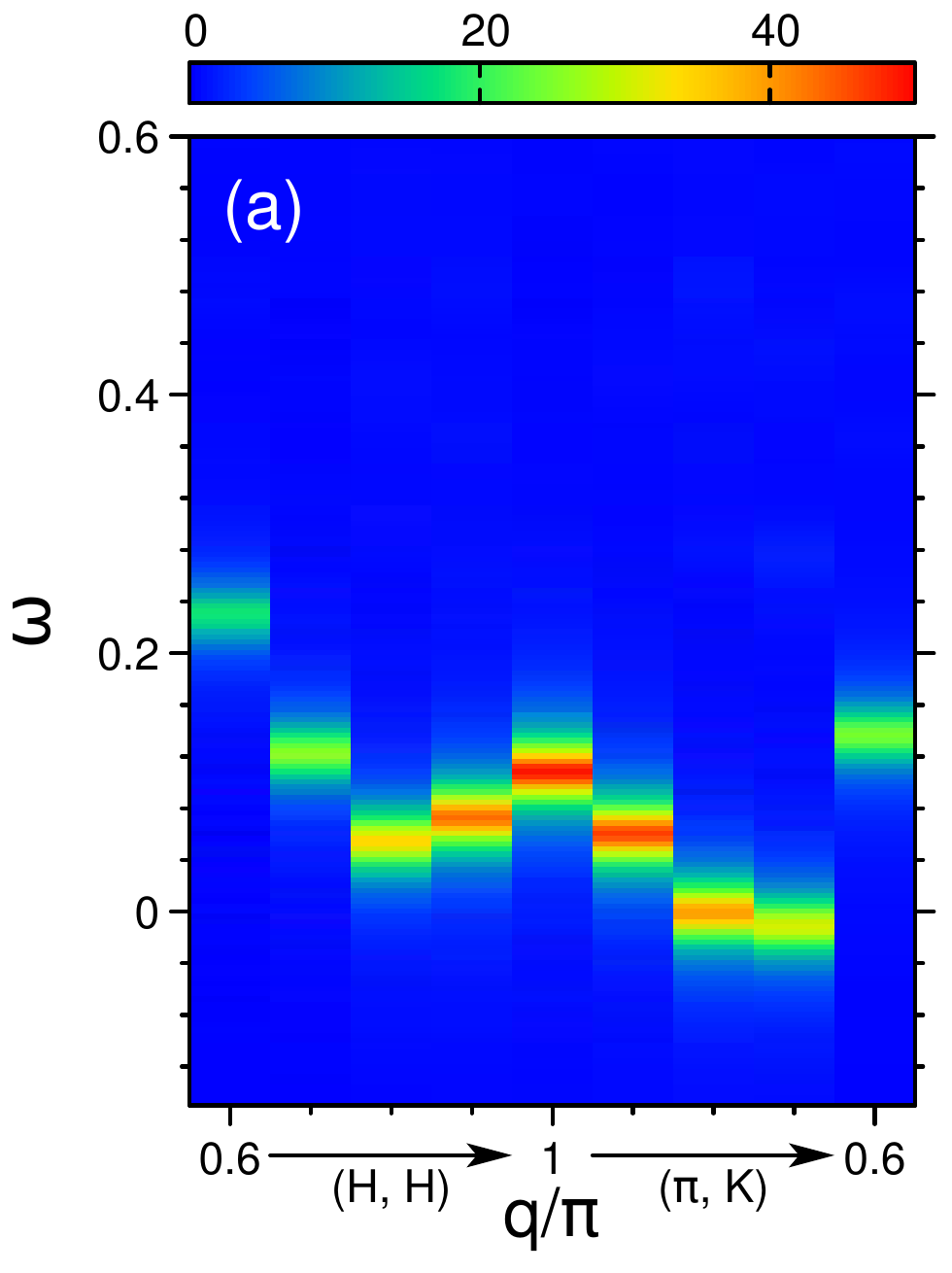}
	\includegraphics[width=0.45\columnwidth,clip=true,angle=0]{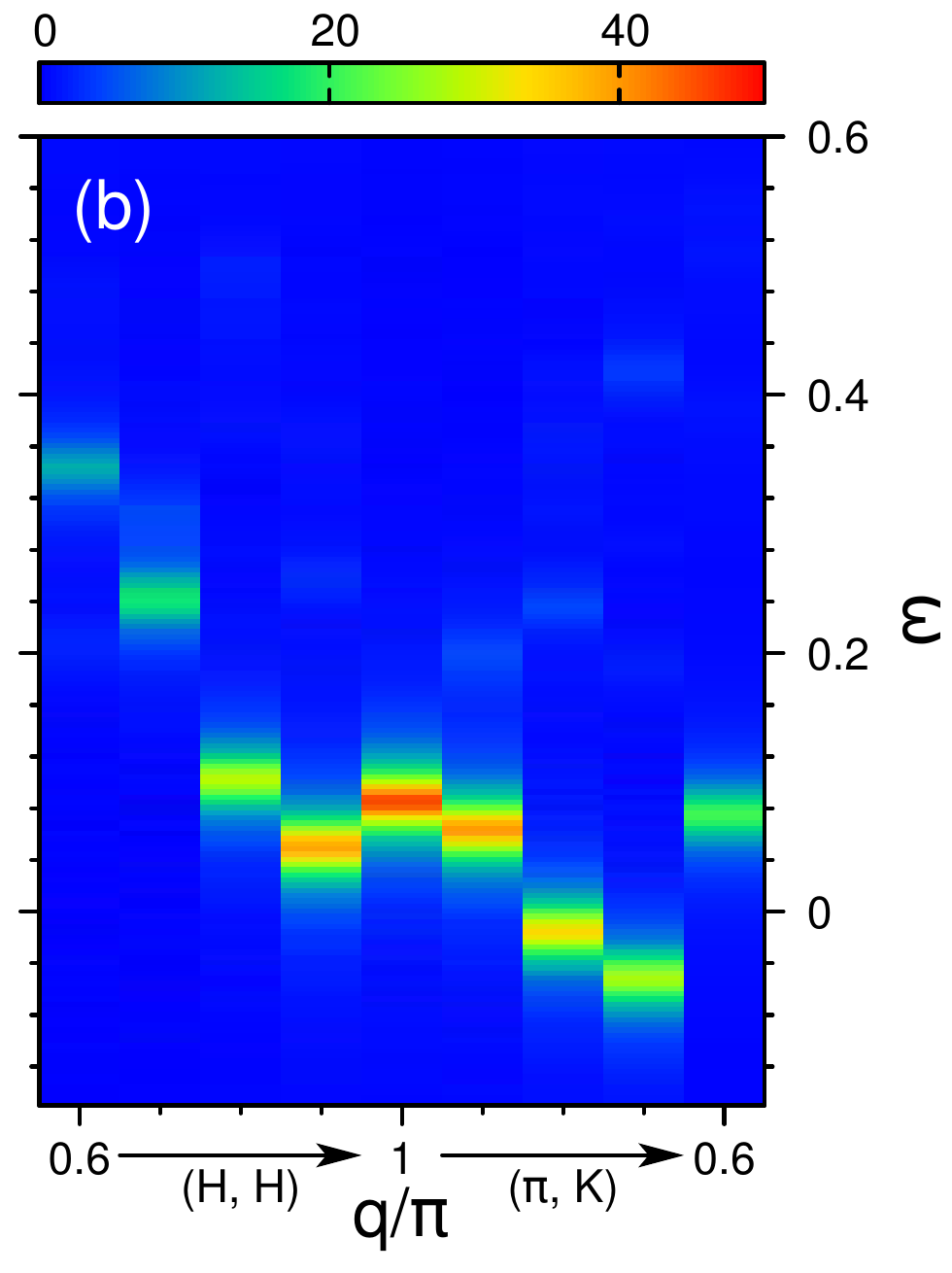}
	\caption{(Color online) $\Im\chi^{-+}(q,\omega)$ along high-symmetric line for (a) RVB state and (b) DDW state. The direction is from diagonal $(H,H)$ to vertical $(\pi, K)$.}
\label{F1}
\end{figure}

RVB state is obtained by projecting BCS mean-field state into the no double-occupancy space. We study the doping level $x=0.13$. The resultant energy is $E=-0.432(8)$ with the optimized variational parameters $\Delta=0.20$, $\mu=-0.78$, and $dt'=-0.01$. The transverse spin susceptibility $\Im\chi^{-+}(q,\omega)$ as function of momentum $q$ and energy $\omega$ is shown in Fig.~\ref{F1}(a). The maximum intensity of the spectrum locates at $Q=(\pi, \pi)$ with the energy about $0.1$, i.e., the resonance energy. It is consistent with the experimental data observed in most moderately underdoped cuprates\cite{fujita2011} and the previous numerical data using similar VMC method\cite{li2010}. Below the resonance energy, the strong intensity can be found both along the diagonal $(H,H)$ and the vertical $(\pi,K)$ directions, though the latter has the slightly stronger intensity and lower peak energy at $q\sim0.8\pi$. Therefore, the lower branch of the hour-glass shaped magnetic excitations can be qualitatively established within the RVB framework. However, no visible intensity above the resonance energy is found in the RVB framework. Beside the lower branch of the hour-glass shaped magnetic dispersion, an outward dispersion is evident which is similar to the spin-wave description in the local picture and may be related to the residue of the strong antiferromagnetic background.

The projected DDW state is exactly the same as RVB state at half-filling due to $SU(2)$ gauge symmetry\cite{zhang1988_1}. Although it has relatively higher energy upon doping, it has been long believed that DDW is the hidden order in the pseudogap state of cuprates\cite{chakravarty2001,laughlin2014} because Hamiltonian that stabilizes the d-wave superconductivity certainly stabilizes the DDW\cite{laughlin2014_1}. The competition between the DDW and superconductivity generates the back-bending behavior of the characteristic temperature of pseudogap under the superconducting dome\cite{yu2017}, in agreement with the recent ARPES measurements\cite{vishik2012,hashimoto2015} and providing the simple explanation on the anomalous thermal evolution in cuprates\cite{kaminski2015,guyard2008}.
In our previous work, we obtained an effective Hamiltonian similar to the DDW mean-field Hamiltonian after taking into account the effects of strong correlation and antiferromagnetic background\cite{zhou2013}. The hour-glass shaped magnetic excitation is well reproduced in the DDW state under the RPA theory, providing a non-d-wave superconductivity origin\cite{zhang2013}. Here we consider DDW as a putative state to study its spin excitations. The only variational parameter is DDW order $D$ which gives the ground state energy $E=-0.424(7)$ with $D=0.20$. The spectrum of magnetic excitations in the DDW state exhibit the negligible difference from that in the RVB state as shown in Fig.~\ref{F1}(b), though very weak intensity is found above the resonance energy.

It is noteworthy that the dispersion of the magnetic excitations along the vertical direction extends to the slightly negative energy near $(\pi,0.7\pi)$ in both the RVB and DDW states, similar to the previous numerical results\cite{li2010,dalla_piazza2014}. This indicates that the RVB and DDW trial states are not the true ground state of the $t$-$J$ model. The corresponding energy at $(\pi,0.7\pi)$ seems to be more negative in the DDW state (Fig.~\ref{F1}(b)), which has a higher total energy.

\subsection{Stripe state}
Previous VMC study showed that the stripe state has lower energy than the uniform RVB state near $1/8$ doping in presence of the next-neighbor hopping\cite{himeda2002}. Further study including the density Jastrow projection revealed a considerable small energy difference between the two states\cite{chou2008}. More recently, the study of the $t$-$J$ model based on the tensor network algorithm also showed the two states are energetically comparable\cite{corboz2014}. From the theoretical perspective at this stage, it is hard to tell whether the ground state of the $t$-$J$ type model is stripe or not.

\begin{figure}[ht]
	\centering
	\includegraphics[width=0.8\columnwidth]{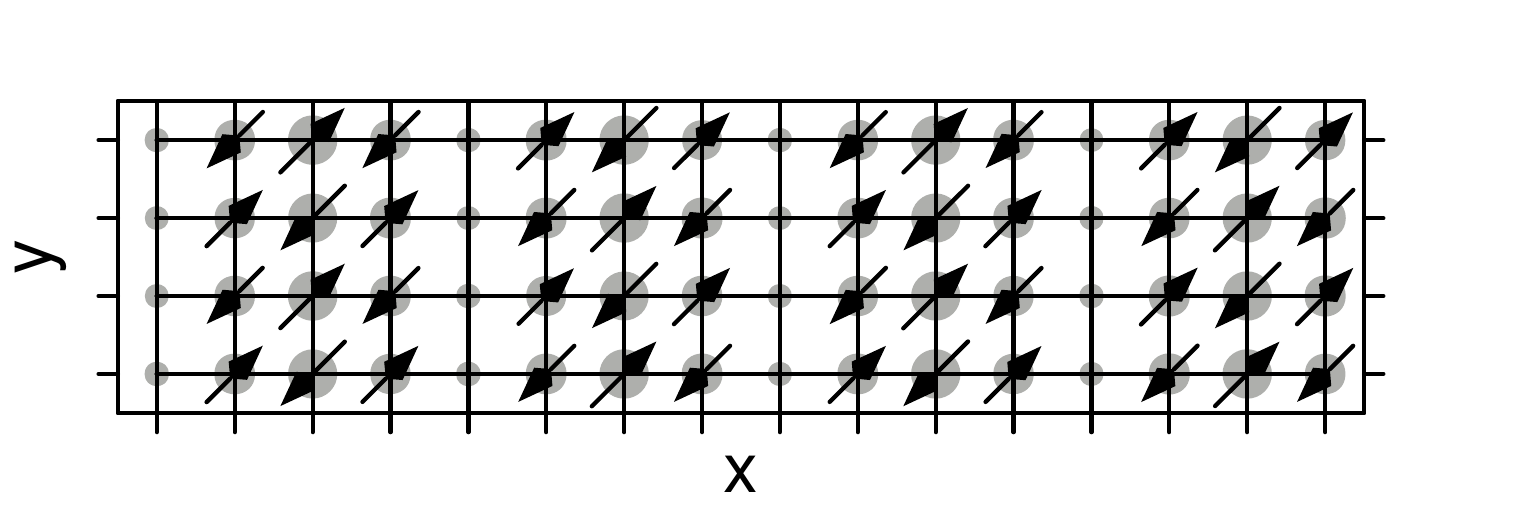}
	\caption{(Color online) Distribution of charge and spin density in real space. The size of grey circles indicates the charge density $\sum_{\sigma}\langle c_{i\sigma}^{\dagger}c_{i\sigma}\rangle$. The black arrows denote the spin moment along z direction $1/2\sum_{\sigma}\sigma\langle c_{i\sigma}^{\dagger}c_{i\sigma}\rangle$. }
\label{F2}
\end{figure}

Here we study the $8a$-period stripe state with both the spin and charge density modulation, resembling the La-, Hg- and Bi-based cuprates\cite{tranquada1995,comin2015_1,neto2014}. The size of the matrix $H_{k,k'}$ and $N_{k,k'}$ is much expanded due to the unit cell enlargement in the stripe state. As a consequence, much more Monte Carlo steps are required in order to obtain the meaningful results and thus it is numerically more demanding. We restrict the calculation on $16\times16$ lattice with the antiperiodic boundary condition along the $x$ direction and periodic boundary condition along the y direction, which satisfies the close-shell condition. The explicit form for the charge, and spin order is $m_{i}=\pm m\sin(q\cdot r_{i})$, and $n_{i}=n\cos(2q\cdot r_{i})$ with the variational parameter $m$, $n$, respectively. The variational parameters that minimize the energy are $m=2.43$, $n=-0.20$, $dt'=-0.27$, $\mu=-1.40$ at doping $x=0.125$, giving the ground state energy $E=-0.433(5)$. The modulation of the charge and spin orders in the real space is schematically shown in Fig.~\ref{F2}. The maximum magnetic moment is almost fully polarized in this state. The spin modulations along the x-direction are doubly periodic compared with the charge modulations since the spin order is anti-phased between the adjacent stripes.

\begin{figure}[ht]
	\centering
	\includegraphics[width=0.45\columnwidth,clip=true,angle=0]{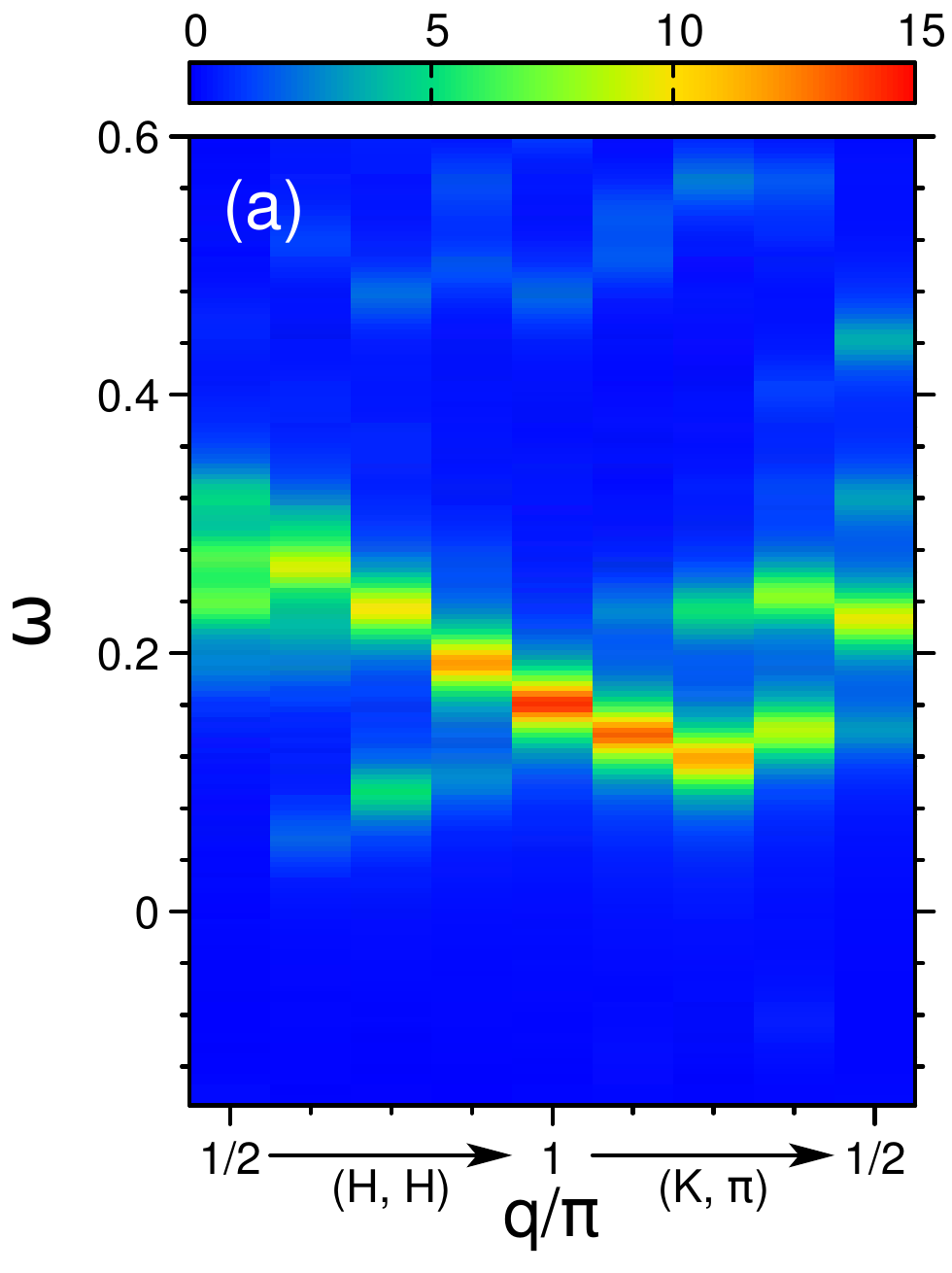}
	\includegraphics[width=0.45\columnwidth,clip=true,angle=0]{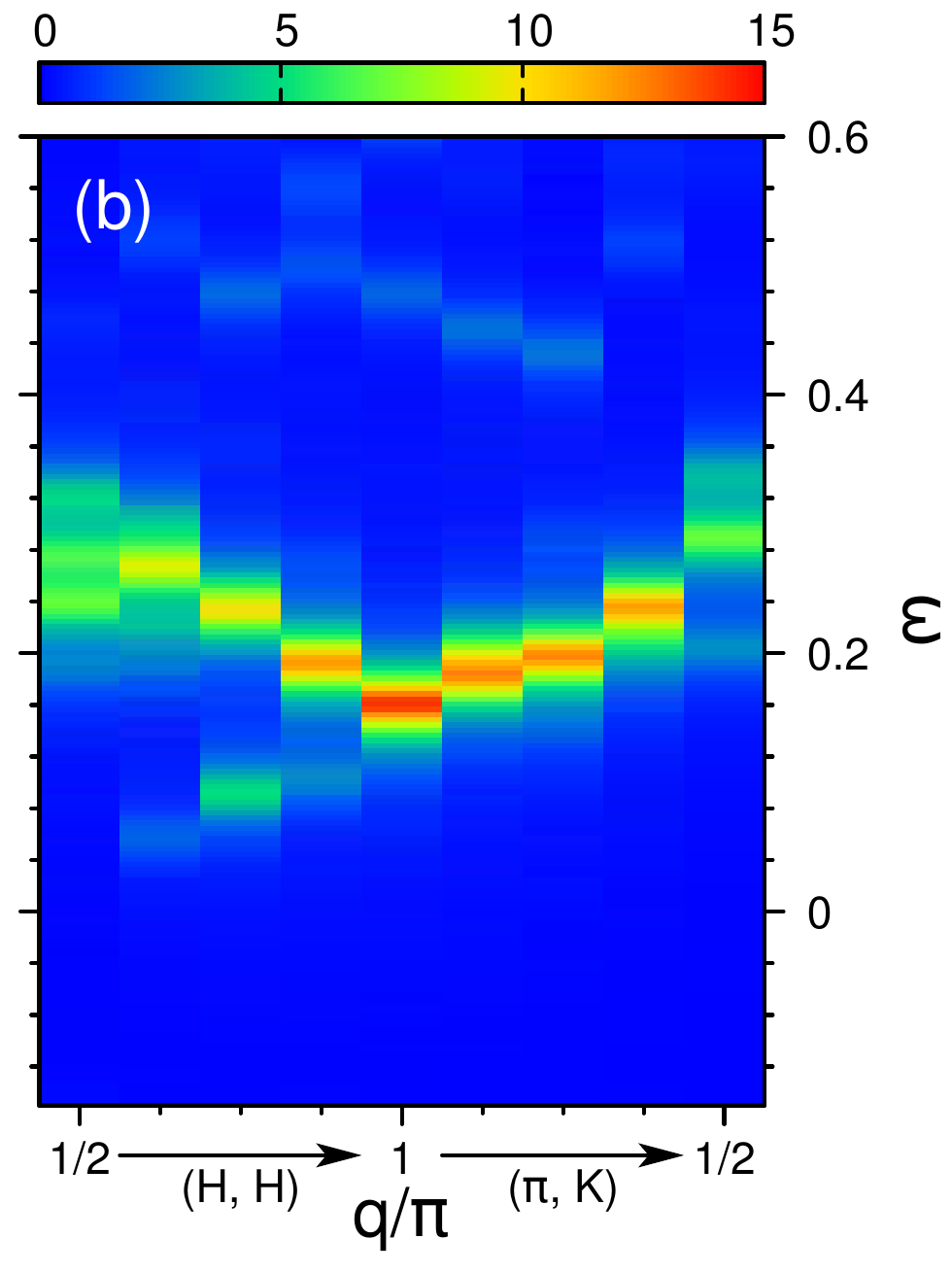}
	\caption{(Color online) $\Im\chi^{-+}(q,\omega)$ for stripe state along high-symmetric line from (a) $(H,H)$ to $(K, \pi)$ and (b) $(H,H)$ to $(\pi, K)$.}
\label{F3}
\end{figure}

Fig.~\ref{F3}(a) shows the spin excitations in the $1/8$-doped stripe state. The resonance at $(\pi, \pi)$ is still present but with slightly weakened intensity. Compared with the above mentioned RVB and DDW state, the outward dispersion is almost invisible with its intensity transferring to the higher energy, forming the upper branch of the hour-glass dispersion. In addition, the lower inward dispersion of the hour-glass remains but with much reduced intensity along the diagonal direction, in agreement with the structural transition from the vertical direction in the lower branch to the diagonal direction in the upper branch. This feature can also be found in some spin-wave treatments as well as RPA based Gutzwiller approximation\cite{kruger2003,carlson2004,seibold2006}, where the weight transfers from the low-energy acoustic branch to the high-energy optical branch as one goes away from $(\pi, \pi)$. Due to the C$_4$ rotational symmetry breaking in the stripe state, the data along the $(\pi,K)$ direction is also shown in Fig.~\ref{F3}(b) for comparison, where the upper branch exists only. In any case, both the lower and upper branch of the hour-glass dispersion are well reproduced in the stripe state.

\begin{figure}[ht]
	\centering
	\includegraphics[width=0.45\columnwidth,clip=true,angle=0]{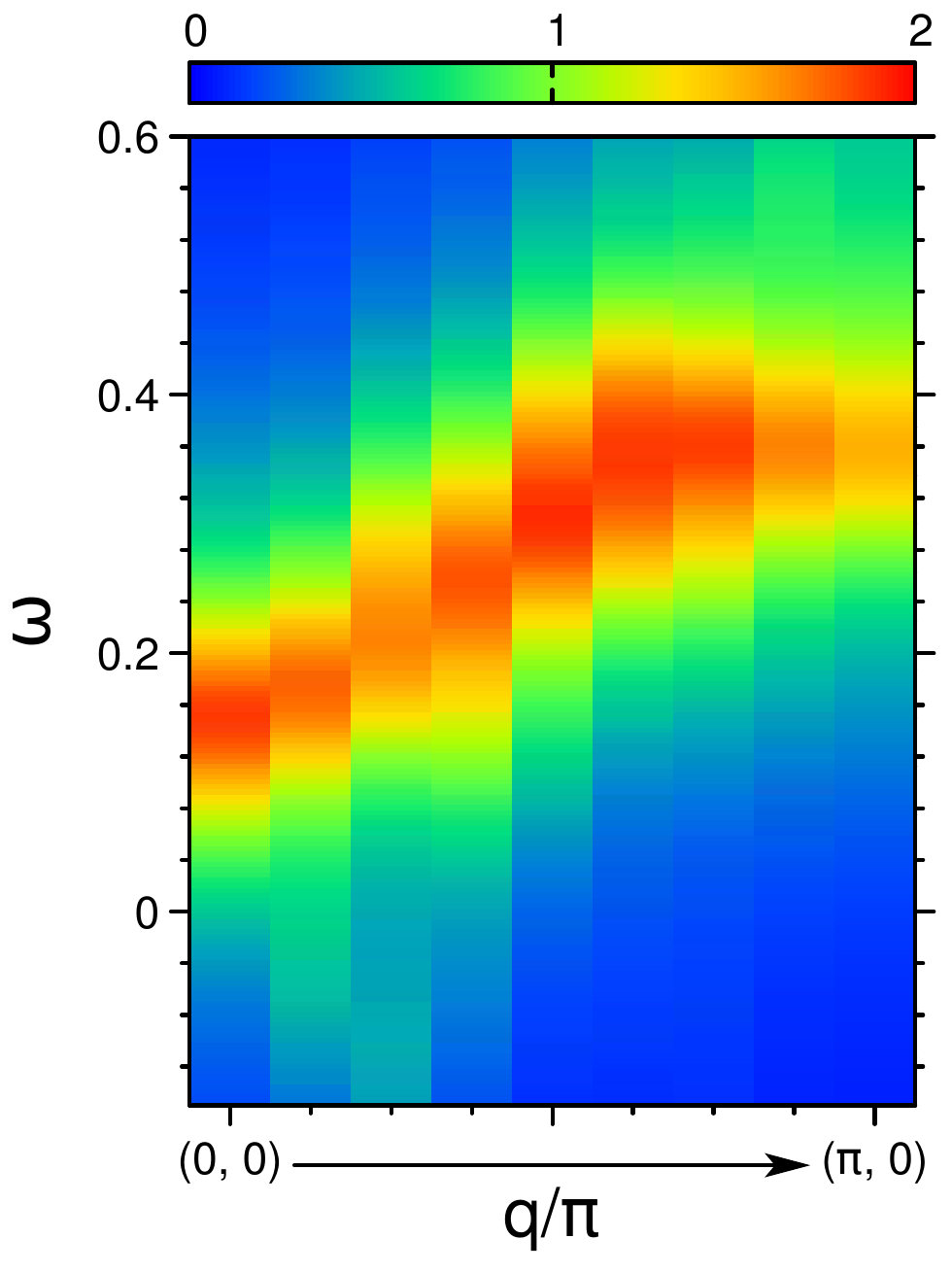}
	\caption{(Color online) $\Im\chi^{-+}(q,\omega)$ for stripe state along $(0,0)$ to $(\pi, 0)$ with the broadening factor $\Gamma=0.1$.}
\label{F3_1}
\end{figure}

As mentioned above, the hour-glass shaped spin dynamics is perhaps not directly related to the superconductivity, but rather a universal feature of the normal state. Recent experimental progresses provide strong evidence of the tendency of the charge order in various types of cuprates\cite{wu2012,wu2011,parker2010,comin2015_1,neto2014,da_silva_neto2015}, especially the potential stripe order\cite{tranquada1995,comin2015}. Similar hour-glass feature is also discovered in the stripe ordered La$_{5/3}$Sr$_{1/3}$CoO$_{4}$\cite{boothroyd2011} and La$_{2-x}$Sr$_{x}$NiO$_{4}$\cite{woo2005}, where no superconductivity is detected, providing the compelling evidence of the stripe origin. The upper branch of the dispersion more likely originates from the band folding caused by the translational symmetry breaking in the stripe state. Our VMC results therefore indicate that the hour-glass feature in the spin dynamics of cuprates is a direct consequence of the stripe state. For completeness, we have checked the magnetic excitations along the $(0,0)$ to $(\pi,0)$ direction as shown in Fig.~\ref{F3_1}. The paramagnon-like excitations observed by the resonant inelastic X-ray scattering\cite{le_tacon2011} is also reproduced in the stripe state.

The above mentioned negative energy found in the RVB and DDW state does not emerge in the stripe state, which has the lowest energy and is considered as the possible origin of the hour-glass feature. Therefore, the absence of negative energy spectrum in the stripe state further supports that the hour-glass feature of the spin dynamics is related to the stripe state.

\subsection{Discussion}

\begin{figure}[ht]
	\centering
	\includegraphics[width=0.45\columnwidth,clip=true,angle=0]{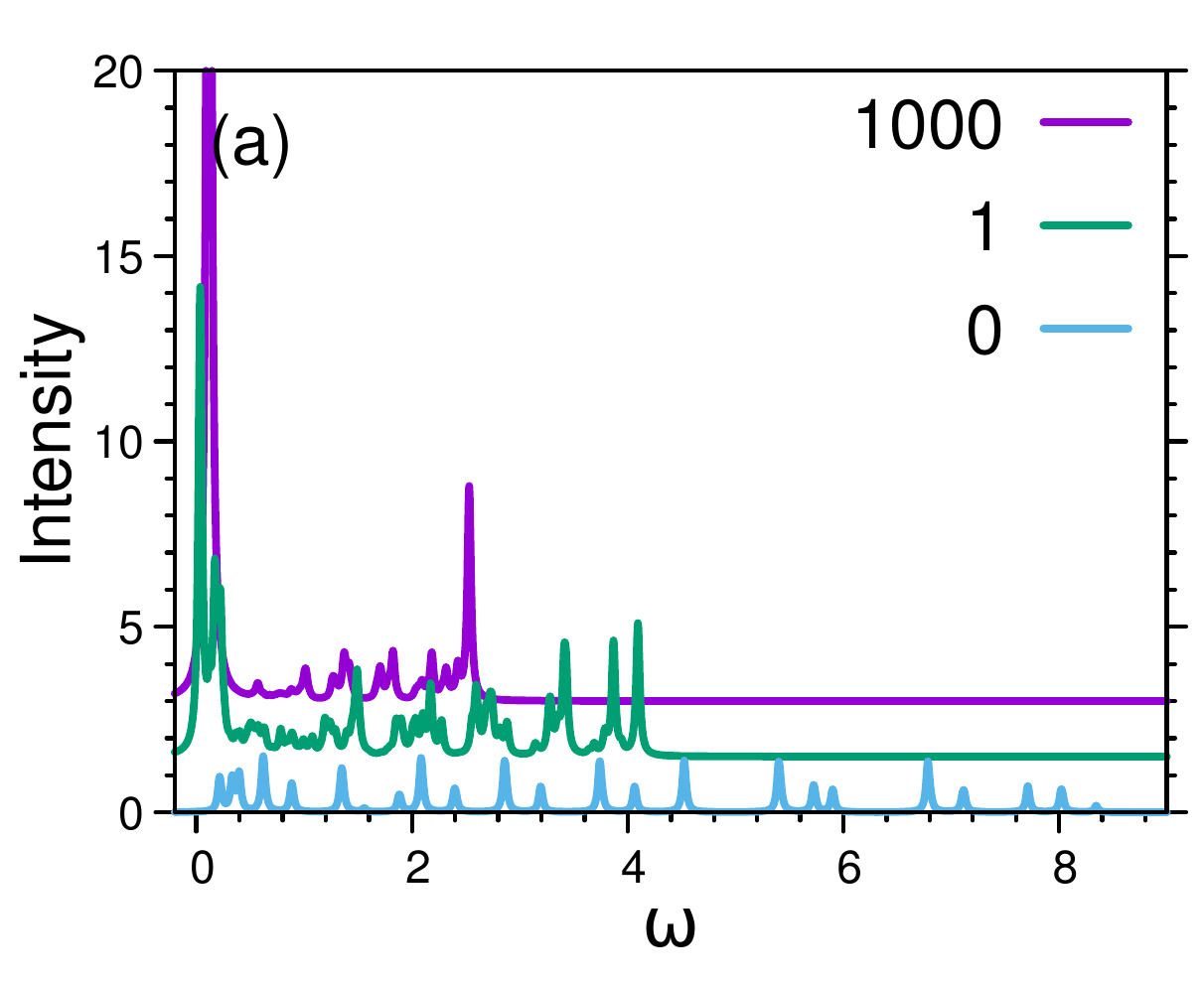}
	\includegraphics[width=0.45\columnwidth,clip=true,angle=0]{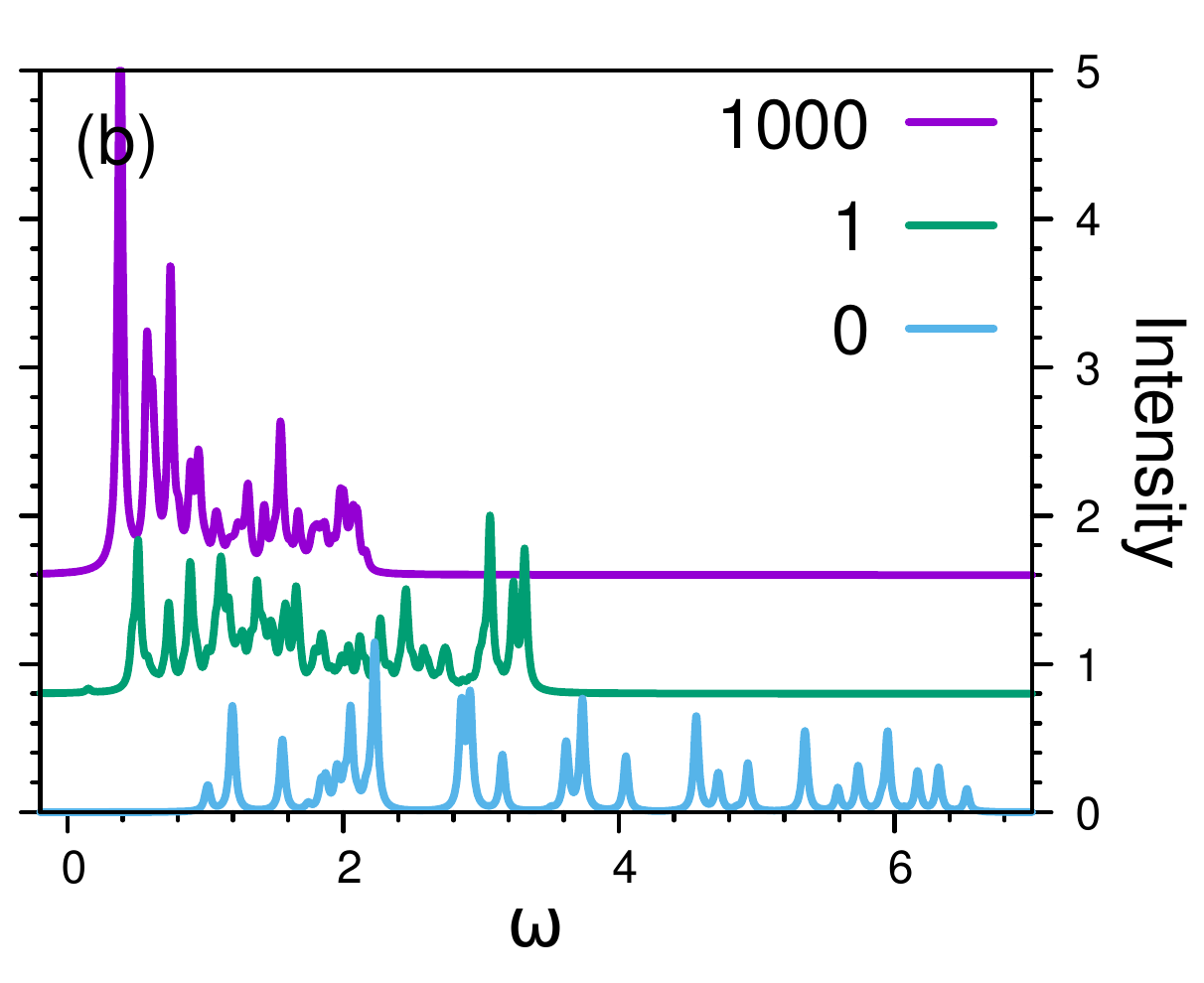}
	\caption{(Color online) Value of $\Im\chi^{-+}(q,\omega)$ for different $\alpha$. (a) $q=(\pi, \pi)$ and (b) $q=(\pi, 0)$ for $\alpha=$0, 1 and 1000. The line of different $\alpha$ is shift in the vertical direction to better distinguish with each other.}
	\label{F5}
\end{figure}

The spin dynamics obtained by the VMC is in sharp contrast to the RPA treatment, where the upper branch of hour-glass dispersion is evident even in the RVB and DDW state. The essence is that the no double-occupancy constraint in spin correlations is treated exactly in the present VMC calculation. To understanding the role of the constraint, we naively release the Gutzwiller projection $P_{G}$ into Jastrow projection $P_{J}=e^{-\alpha \sum_{i,j}n_{i}n_{j}}$ with $i=j$.  Fig.~\ref{F5}(a) shows the energy dependence of $\Im\chi^{-+}(q,\omega)$ at the commensurate position $(\pi,\pi)$ in the RVB state. The spectrum is a continuum in the non-projected limit $\alpha\rightarrow 0$ and develops into sharp peak at the resonance energy in the Gutzwiller limit $\alpha\rightarrow \infty$. Similar results are also found in the DDW state. In this sense, the resonance is a consequence of the no double-occupancy constraint. This may be further argued in the overdoped cuprates where the resonance peak damps into particle-hole continuum\cite{li2010} due to the weakened constraint of no double-occupancy. In contrast, the spectrum at $(\pi,0)$ remains continuum even in the Gutzwiller limit (see Fig.~\ref{F5}(b)), in agreement with the fractional spin found in the Heisenberg model\cite{dalla_piazza2014}.

\begin{figure}[ht]
	\centering
	\includegraphics[width=0.45\columnwidth,clip=true,angle=0]{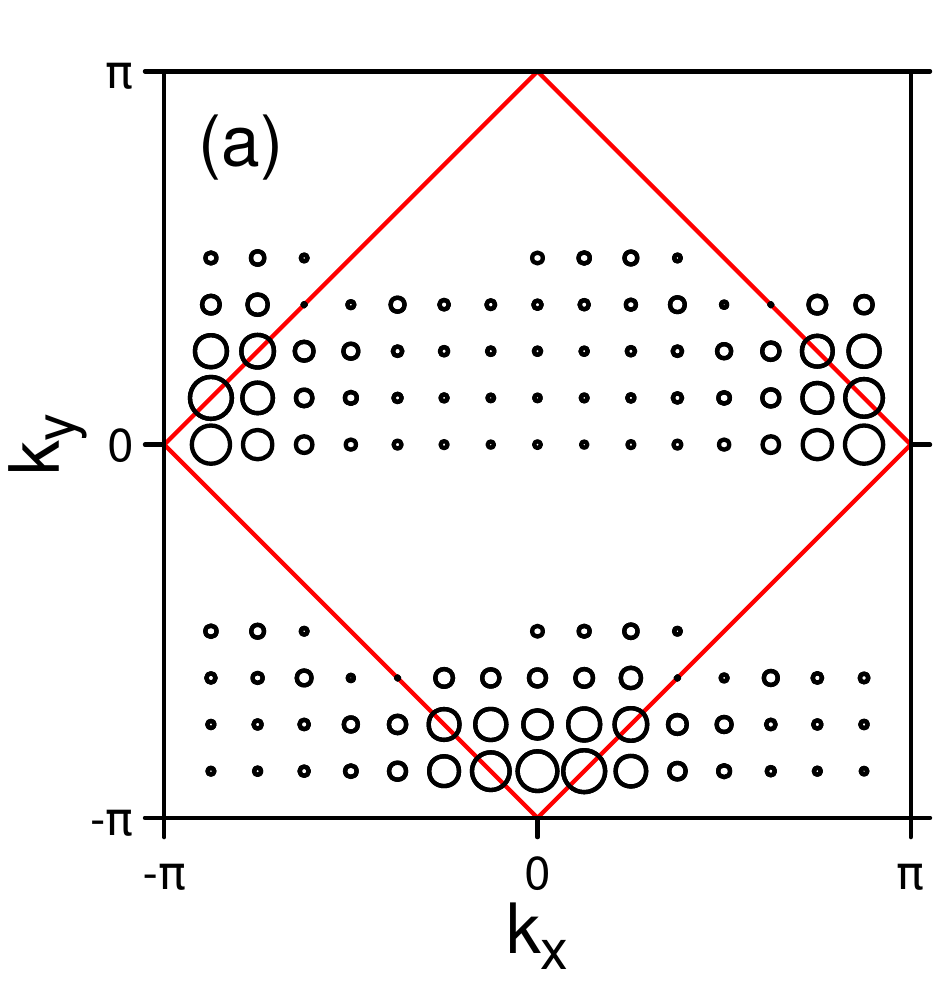}
	\includegraphics[width=0.45\columnwidth,clip=true,angle=0]{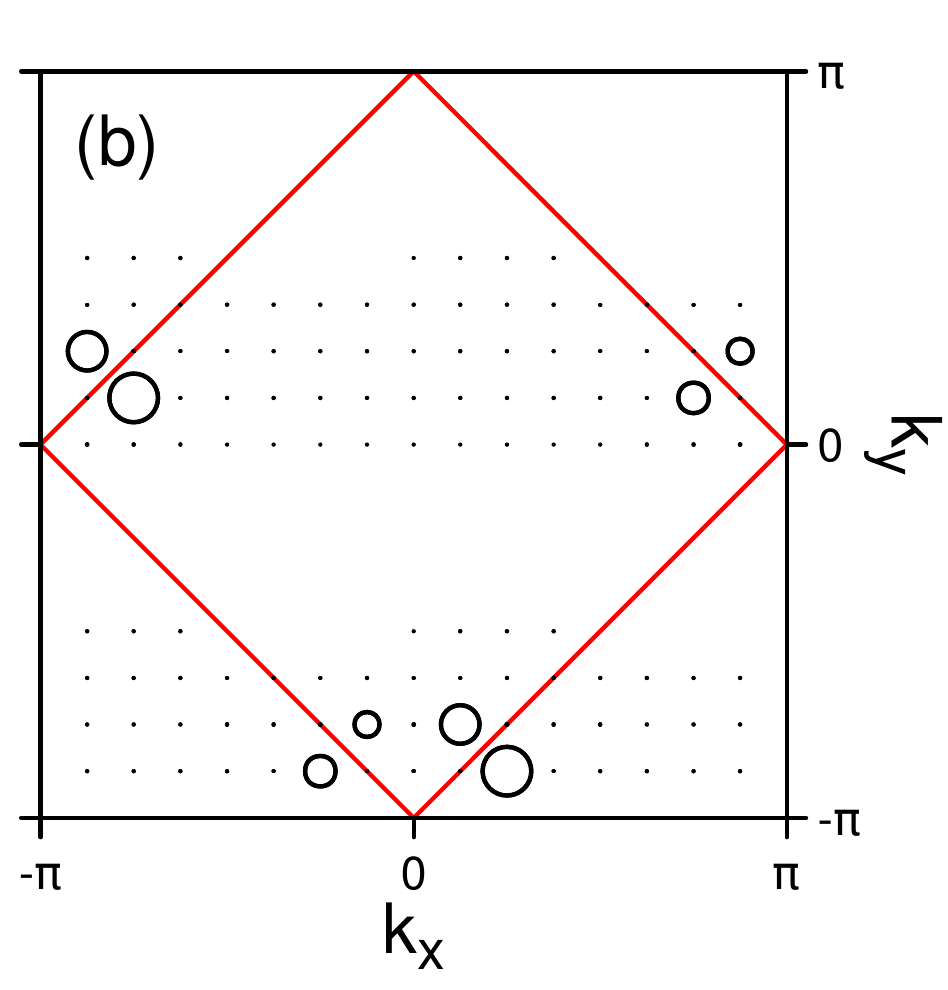}
\caption{(Color online) Distributions of $M(k,\omega)$ in Brillouin zone for $q=(\pi, \pi)$. (a) $M(k,\omega)$ at the resonance energy in the fully projected RVB state. (b) the same but at energy of the maximum intensity in the non-projected RVB state. The relative value is denoted by cycle size.}
\label{F6}
\end{figure}

\begin{figure}[ht]
	\centering
	\includegraphics[height=0.45\columnwidth,clip=true,angle=0]{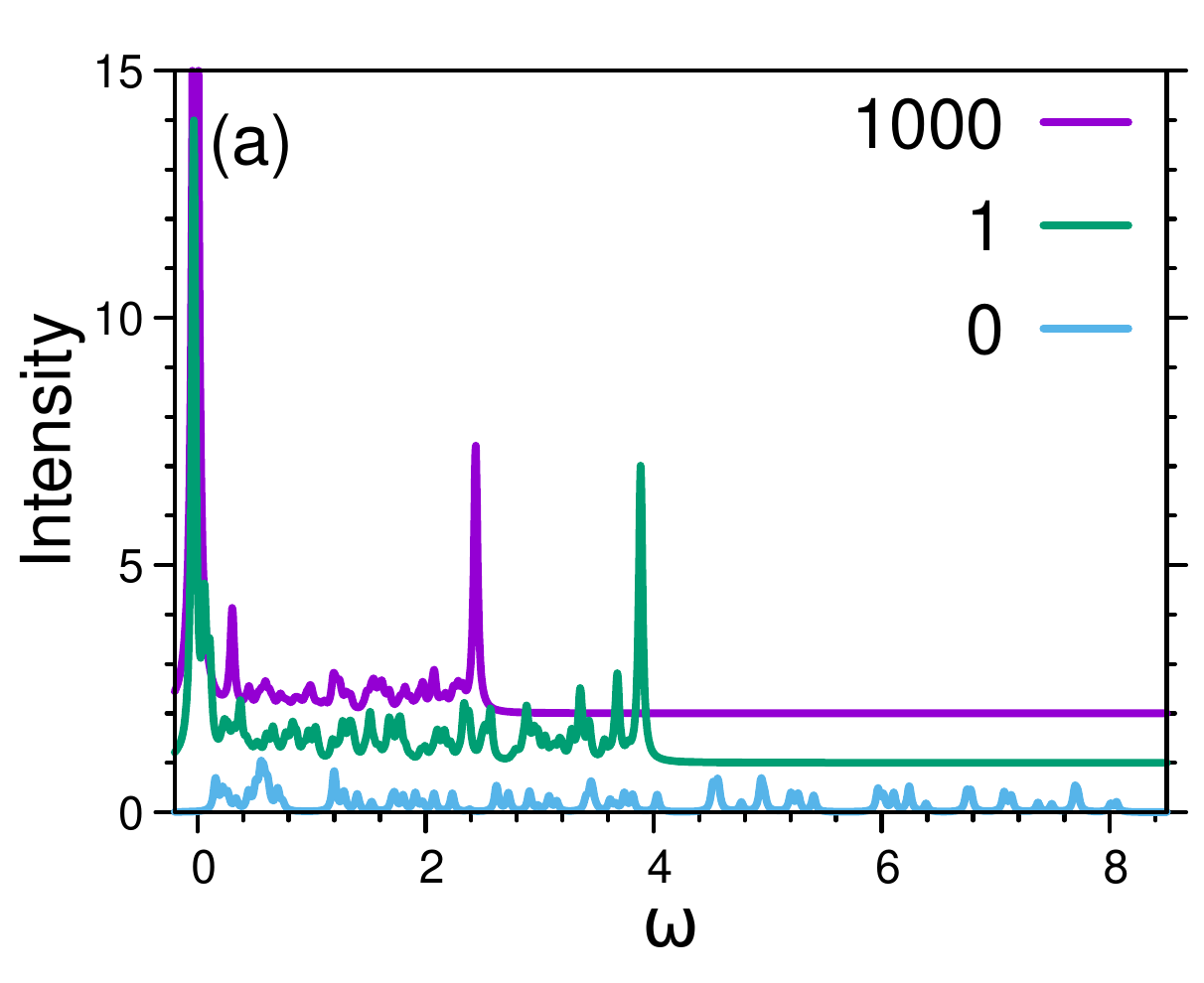}
	\includegraphics[height=0.45\columnwidth,clip=true,angle=0]{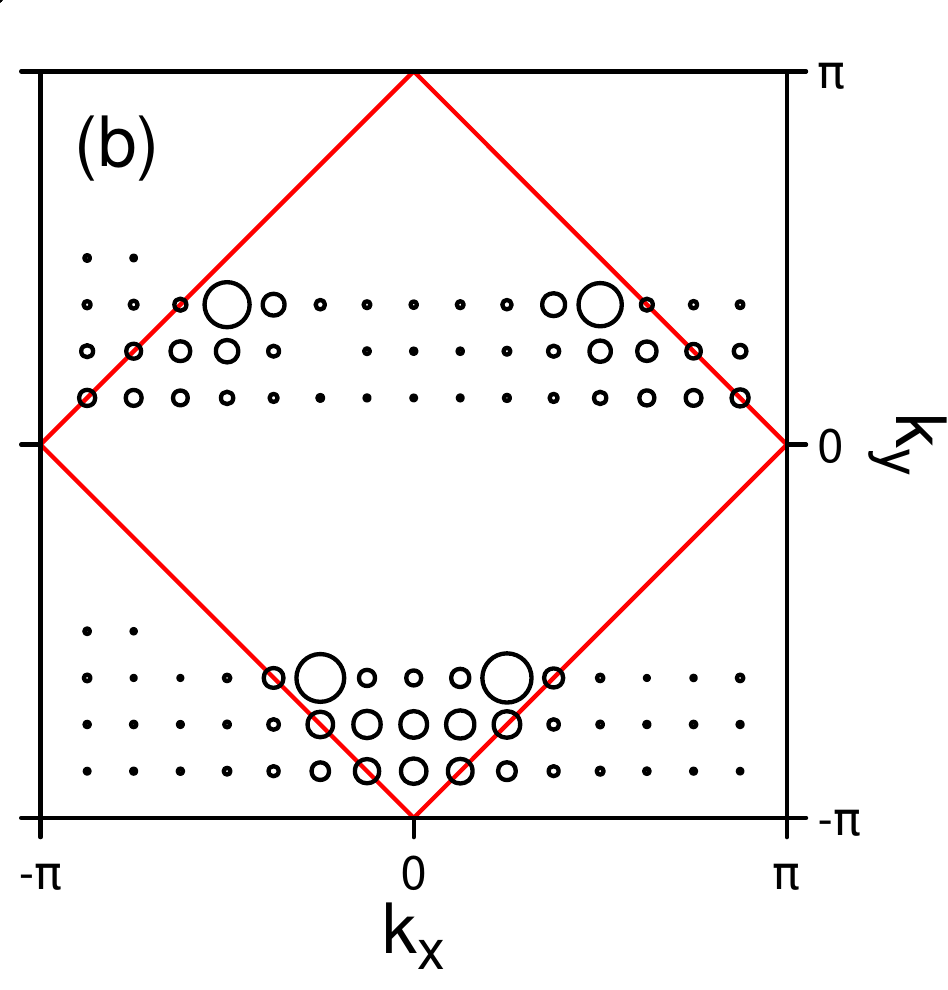}
\caption{(Color online)  (a) Value of $\Im\chi^{-+}(q,\omega)$ for different $\alpha$ at $q=(3/4\pi, \pi)$. (b) Distributions of $M(k,\omega)$ in Brillouin zone for $q=(3/4\pi,\pi)$ at the maximum intensity of the RVB state. The relative value is denoted by cycle size.}
\label{F6p}
\end{figure}

To gain further insight into the excitation spectrum, we define $M(k,\omega)=\sum_{n}\vert\langle k\vert n\rangle \langle n|S_{q}^{+}\vert G\rangle\vert^{2}\delta(\omega-E_{n}+E_{g})$, which roughly extracts the contribution from different excited state $\vert k\rangle$ at given energy $\omega$. Multiple $\vert k\rangle$ states contribute to the intensity at the resonance energy at commensurate position $q=(\pi,\pi)$ in the Gutzwiller limit (Fig.~\ref{F6}(a)), manifesting the resonance is a collective mode, in agreement with the previous statements\cite{brinckmann1999,li2002,onufrieva2002,eremin2005}. In contrast, few $k$ point contributes to the intensity in the non-projected limit (Fig.~\ref{F6}(b)). This strongly indicates that the constraint of no double-occupancy makes the magnetic excitation more collective and thus more resonant. This statement is also valid for $q$ deviating from $(\pi,\pi)$ as shown in Fig.~\ref{F6p}, where the collective nature remains. In some RPA calculations\cite{brinckmann1999,li2002}, the low energy incommensurability comes from the particle-hole excitations, and the resonance is spin exciton lying below the particle-hole continuum. Whereas, others argued that the lower branch of the hour-glass dispersion is also a collective mode, the same as the resonance\cite{onufrieva2002,eremin2005}. Our results therefore support the latter statement.

\section{Summary}
\label{s3}
In summary, we have studied the spin excitations in the $t$-$J$ model using the variational Monte Carlo method. The constructed excited states respect the no double-occupancy constraint and satisfy the local spin sum rule. The parameter-free results of the spin dynamics provide a better evaluation on some potential mean-field trial states for the cuprates. The lower branch of the hour-glass dispersion of the spin excitations is also a collective mode, similar to the resonance. The upper branch is established in the stripe state only. We conclude that the so-called hour-glass feature of the spin dynamics discovered in the cuprates is related to the stripe state, which is found to be universal in the cuprates.

\section*{Acknowledgements}

We thank W.-G. Yin, H.-Q. Lin and J.-X. Li for helpful discussions and suggestions. This work was supported by the NSFC of China Grants No.11274276 and the Ministry of Science and Technology of China 2016YFA0300401. Y. Z. acknowledges the visiting scholarship of Brookhaven National Laboratory and the financial support of China Scholarship Council.

\bibliography{ref}

\end{document}